\pdfoutput=1

\documentclass[11pt,twoside,a4paper,cmspaper,final,collab]{cms-tdr}
\def\svnVersion{e791ba0-D}\def\svnDate{2020/01/02}\def\cmsCernNoTag{CERN-EP-2019-167}\def\cmsCernDate{\today}\def\cmsMessage{"Published in the European Physical Journal C as \href{http://dx.doi.org/10.1140/epjc/s10052-019-7541-6}{\doi{10.1140/epjc/s10052-019-7541-6}.}"}

\begin{document}\cmsNoteHeader{SMP-18-015}

\hyphenation{had-ron-i-za-tion}
\hyphenation{cal-or-i-me-ter}
\hyphenation{de-vices}

\cmsNoteHeader{AN-18-065}

\title{Evidence for $\PW\PW$ production from double-parton interactions in proton-proton collisions at $\sqrt{s} = 13$\TeV}

\date{\today}

\newlength\cmsTabSkip\setlength{\cmsTabSkip}{1ex}
\newcommand{\model}{\ensuremath{\,\text{(model)}}\xspace}
\newcommand{\ptvecmissO}{\ensuremath\smash[b]{{\ptvec^{\kern1pt\text{miss\,(1)}}}}\xspace}
\newcommand{\ptvecmissT}{\ensuremath\smash[b]{{\ptvec^{\kern1pt\text{miss\,(2)}}}}\xspace}
\newcommand{\mttwo}{\ensuremath{\mTii(\ell_{1},\,\ell_{2})}\xspace}

\abstract{
A search for $\PW\PW$ production from double-parton scattering processes using same-charge electron-muon and dimuon events is reported, based on proton-proton collision data collected at a center-of-mass energy of 13\TeV. The analyzed data set corresponds to an integrated luminosity of 77.4\fbinv, collected using the CMS detector at the LHC in 2016 and 2017.
Multivariate classifiers are used to discriminate between the signal and the dominant background processes. A maximum likelihood fit is performed to extract the signal cross section. This leads to the first evidence for $\PW\PW$ production via double-parton scattering, with a significance of 3.9 standard deviations. The measured inclusive cross section is $1.41\pm 0.28\stat\pm 0.28\syst$\unit{pb}.}

\hypersetup{%
pdfauthor={CMS Collaboration},%
pdftitle={Evidence for WW production from double-parton interactions in proton-proton collisions at sqrt(s) = 13 TeV},%
pdfsubject={CMS},%
pdfkeywords={CMS, physics, WW production}}

\maketitle

\section{Introduction}
Events in which two hard parton-parton interactions occur within a single proton-proton ($\Pp\Pp$)
collision---referred to as double-parton scattering (DPS) processes---have been discussed theoretically since
the introduction of the parton model~\cite{Goebel:1979mi,Shelest:1982dg,Sjostrand:1987su,dps2,dps1,Blok:2013bpa,dpsTheory,Bartalini:2017jkk}.
Experimentally, such processes have
been studied at hadron colliders at different
center-of-mass energies using multiple final states~\cite{sigeff5,sigeff6,cdf:dps,CDF:dps2,sigeff4,sigeff3,LHCb:dps,sigeff2,sigeff1,Aaij:2015wpa,CMS-FSQ-16-005,atlas:dps2018,Aad:2014kba,Abazov:2015fbl}.

The cross section for a single hard scattering (SHS) can be factorized into a term containing the parton
distribution functions (PDFs) and the partonic cross section of the process at hand, but this approach becomes nontrivial
for DPS processes. Although the factorized partonic cross sections remain unchanged, the PDF term in the DPS case contains
elements from two distinct partons in each proton. This term includes
a distance parameter between the partons in the plane transverse to the direction of motion of each proton.
Precise calculations
of the involved dynamics have been carried out for such a case~\cite{dpsTheory}. Assuming that both
the partonic cross sections and  the transverse and longitudinal parts of the
PDF terms factorize, the DPS cross section can be written in a simplified model as
\begin{linenomath}
\begin{equation}
\label{eq:pocketformula}
    \sigma_{\mathrm{AB}}^{\mathrm{DPS}} = \frac{n}{2} \frac{\sigma_{\mathrm{A}} \sigma_{\mathrm{B}}}{\sigma_{\text{eff}}},
\end{equation}
\end{linenomath}
where ``A'' and ``B'' denote the SHS processes, and $\sigma_{\mathrm{A}}$ and $\sigma_{\mathrm{B}}$ are
their respective production cross sections. The factor $n$ is equal to unity if processes
A and B are the same, and is equal to two otherwise. The parameter $\sigma_{\text{eff}}$, the
effective cross section of DPS processes,
is related to the extent of the parton distribution in the plane orthogonal to the direction of motion
of the protons. It was measured at different hadron colliders and center-of-mass energies in a variety of
final-state processes with comparatively large uncertainties (${\approx}30$\%). Its value
ranges between 15 and 26\unit{mb} for processes involving a vector boson~\cite{sigeff2,sigeff1,sigeff4,sigeff7,sigeff3,LHCb:dps,Aad:2014kba,Aaij:2015wpa}. Significantly lower values, down to 2.2\unit{mb}, are measured for processes
involving heavy-flavor production~\cite{Abazov:2015fbl}.

\begin{figure*}[h!]
\centering
\includegraphics[width=0.285\textwidth]{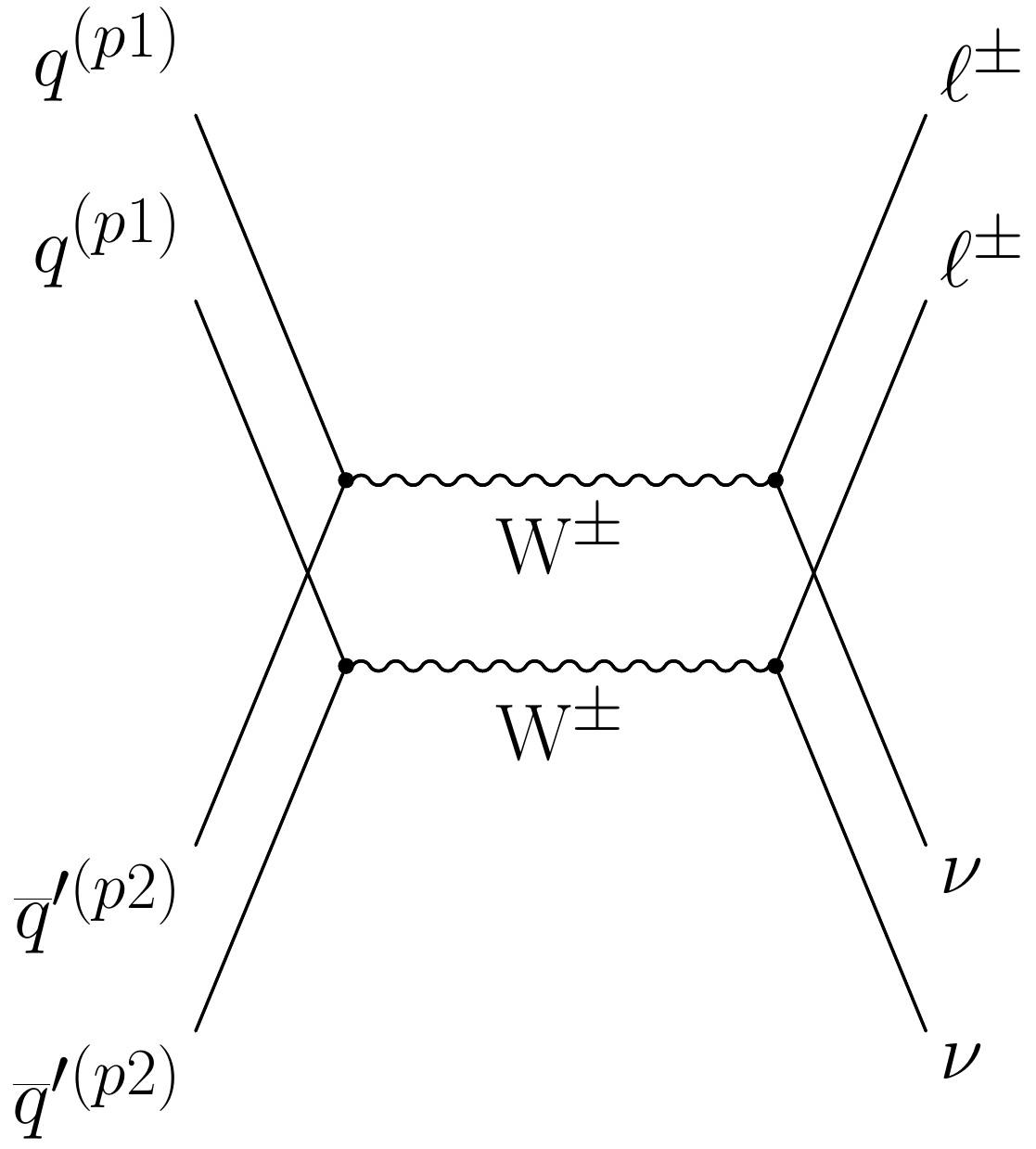}
\includegraphics[width=0.285\textwidth]{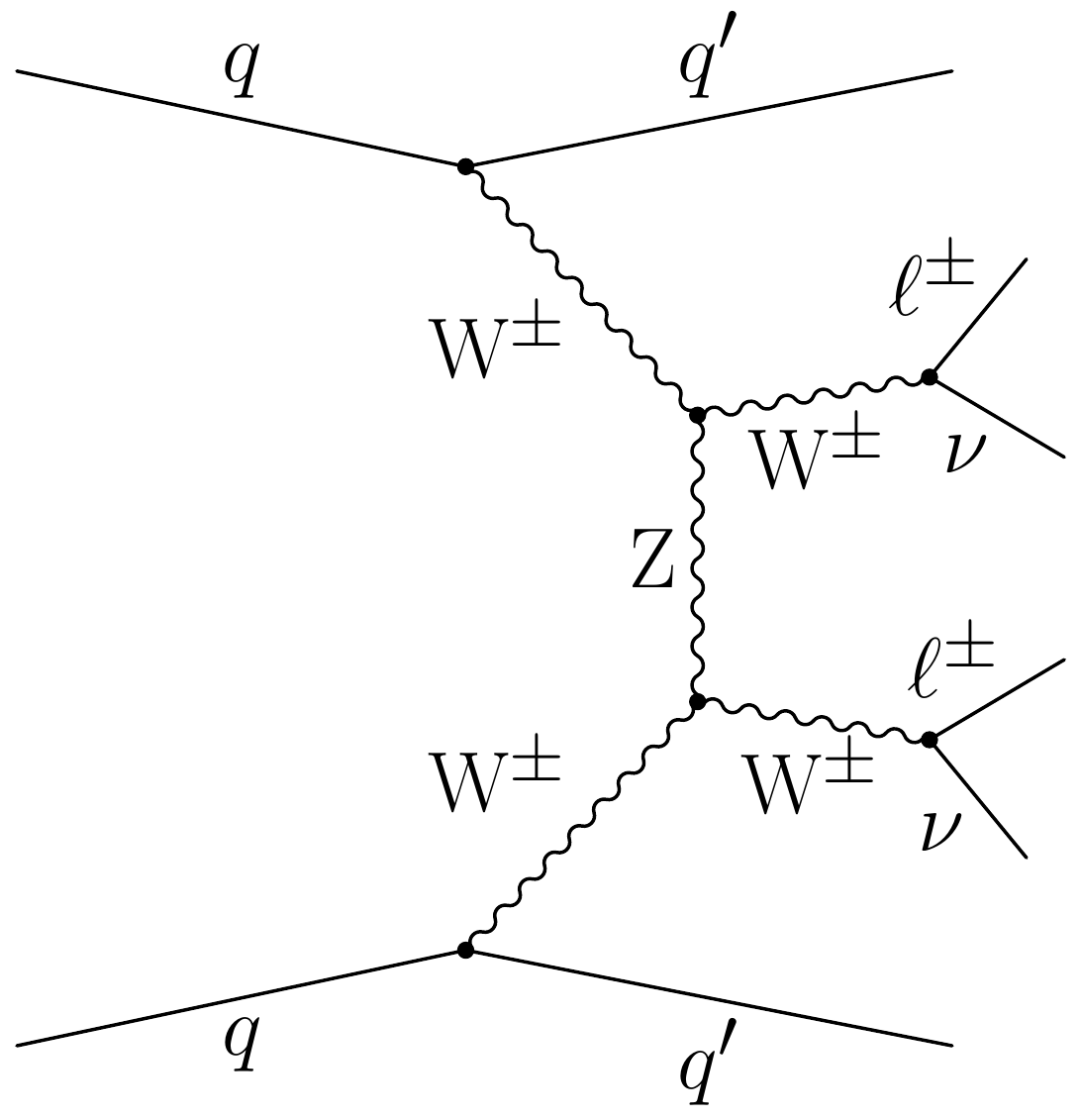}
\includegraphics[width=0.275\textwidth]{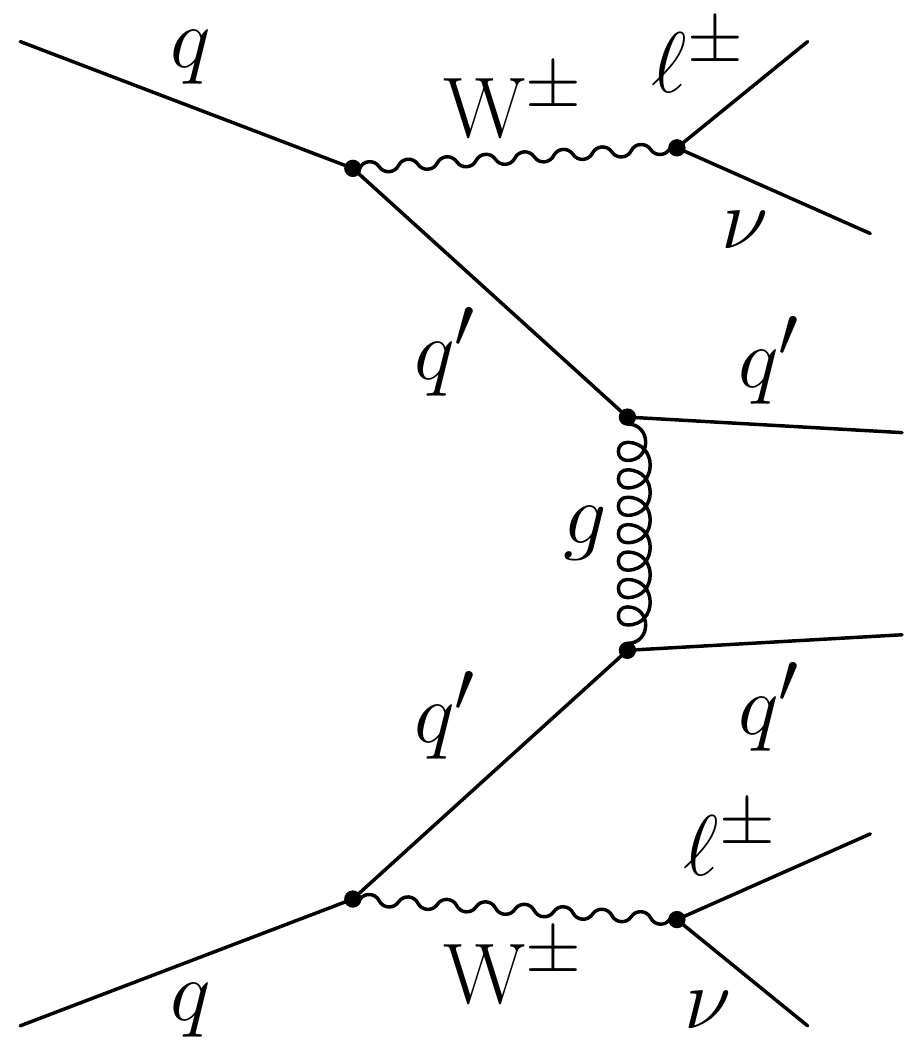}
\caption{Schematic diagrams corresponding to the production of $\PW^{\pm}\PW^{\pm}$ via the DPS process (left) and via SHS  processes (middle and right), with both $\PW$ bosons further decaying leptonically.}
\label{fig:dpsfeyn}
\end{figure*}

One of the most promising processes to study DPS is the case in which both hard scatterings lead
to the production of a $\PW$ boson, and, in particular, the final state with two same-charge $\PW$
bosons~\cite{Kulesza:1999zh}. The SHS $\PW^{\pm}\PW^{\pm}$ production includes two additional partons
and its cross section is therefore suppressed at the matrix-element level.
Figure~\ref{fig:dpsfeyn} illustrates the production of $\PW^{\pm}\PW^{\pm}$
via the DPS process (left) and via SHS processes (middle and right) at leading order (LO) in perturbative
quantum chromodynamics (QCD).
The absence of jets in the $\PW^{\pm}\PW^{\pm}$ production via DPS at
 LO in perturbation theory provides an additional handle to reduce the contributions from the SHS backgrounds
by introducing an upper limit on the number of jets.
Moreover, when both $\PW$ bosons decay leptonically, this event exhibits a clean final state in the
detector, and the excellent reconstruction and resolution of leptons in the CMS detector provides an accurate measurement
of the $\PW\PW$ DPS cross section.

However, DPS $\PW\PW$ production has not been observed experimentally. An observation of this process would permit the
validation of the factorization
approach, which is prevalent in current Monte Carlo (MC) event generators. In addition, it is proposed that angular
observables in the DPS $\PW^{\pm}\PW^{\pm}$ process are sensitive to nontrivial longitudinal momentum correlations
among the partons~\cite{Gaunt,Ceccopieri:2017oqe,Cotogno:2018mfv}. The DPS $\PW^{\pm}\PW^{\pm}$ process also constitutes a background in searches for new physics at the CERN LHC,
\eg, in searches for the electroweak production of supersymmetric particles~\cite{ewino8tev}.
A measurement of the DPS $\PW\PW$ production cross section ($\sigma_{\mathrm{DPS\,\PW\PW}}$) would improve the
reach of such searches.

A search for the production of $\PW^{\pm}\PW^{\pm}$ via DPS was reported in the past by the CMS Collaboration
using $\Pp\Pp$ collisions at $\sqrt{s}$ = 8\TeV, and
an upper limit of 0.32\unit{pb} was set on its production cross section at 95\% confidence level~\cite{CMS-FSQ-16-005}.
An increased production cross section at
$\sqrt{s}$ = 13\TeV and a larger data set collected using the CMS detector allow a
 more detailed study of this rare and interesting physics process. This paper presents a measurement of this process
 performed with $\Pp\Pp$ collision data, recorded using the CMS experiment at a
center-of-mass energy of 13\TeV in 2016 and 2017.
The analyzed data sample corresponds to an integrated luminosity of 77.4\fbinv.

The analysis focuses on the leptonic decay of the $\PW$ bosons in final states consisting of a same-charge
electron-muon ($\Pepm\PGmpm$) or dimuon ($\PGmpm\PGmpm$) pair,
which include small contributions from leptonic $\tau$ decays.
The dielectron final state is not considered because of the relatively higher level of backgrounds.

\section{The CMS detector}
\label{sec:cms}

The central feature of the CMS apparatus is a superconducting solenoid, 13\unit{m} in length and 6\unit{m} in diameter, which
provides
an axial magnetic field of 3.8\unit{T}. The bore of the solenoid is outfitted with various particle detection systems.
Charged-particle trajectories are measured in the silicon pixel and strip trackers, covering $0<\phi<2\pi$ in azimuth and $\abs{\eta}<2.5$,
where the pseudorapidity $\eta$ is defined as $\eta = -\ln [\tan(\theta/2)]$, with $\theta$ being the polar angle of the
trajectory of the particle with respect to the counterclockwise direction. A crystal electromagnetic calorimeter (ECAL),
 and a brass/scintillator
hadron calorimeter surround the tracking volume. In this analysis, the calorimetry provides high resolution energy
and direction measurements of electrons and hadronic jets. A preshower detector consisting of two planes of silicon sensors
interleaved with lead is located in front of the ECAL at $\abs{\eta}>1.479$. Muons are measured in gas detectors embedded in
the steel flux-return yoke outside the solenoid. The detector is nearly hermetic, allowing energy balance measurements in the
plane transverse to the beam directions. A two-tier trigger system selects the most interesting pp collision events for use
in physics analysis~\cite{Khachatryan:2016bia}. A more detailed description of the CMS detector can be found in Ref.~\cite{Chatrchyan:2008zzk}.

\section{Event selection criteria}
\label{sec:objectsamples}
{\tolerance=300
A particle-flow (PF) technique is used to reconstruct and identify particles in the event~\cite{Sirunyan:2017ulk}. It
combines subdetector-level information to reconstruct individual particles and identify them as charged and
neutral hadrons, photons, and leptons. Electron and muon candidates are reconstructed by associating a charged-particle
track reconstructed in the silicon detectors with a cluster of energy in the ECAL~\cite{Khachatryan:2015hwa} or
a track in the muon system~\cite{Sirunyan:2018fpa}. These PF candidates are used to reconstruct higher-level
objects, such as jets, hadronically decaying $\tau$ leptons (\tauh), and missing transverse momentum (\ptmiss).
The missing transverse momentum vector $\ptvecmiss$ is computed as the negative vector \pt sum of
all the PF candidates in an event, and its magnitude is denoted as \ptmiss~\cite{CMS-PAS-JME-17-001}.
The \tauh candidates are reconstructed via the ``hadrons plus strips'' algorithm~\cite{Sirunyan:2018pgf}
and are further selected using a multivariate (MVA) classifier to reduce the misidentification rate of
light-quark and gluon jets.
\par}

{\tolerance=300
The reconstructed vertex with the largest value of
summed physics-object $\pt^2$ is the
primary $\Pp\Pp$ interaction vertex. Jets are reconstructed from charged and neutral PF candidates clustered using the anti-\kt clustering algorithm~\cite{Cacciari:2008gp,FastJet}
with a distance parameter of 0.4, as implemented in
the \FASTJET package~\cite{FastJet,Cacciari:2005hq}.
\par}

Two $\PQb$ tagging algorithms, which depend on the year of data taking~\cite{Chatrchyan:2012jua,Sirunyan:2017ezt}, are
used to identify jets originating from $\PQb$ quarks. They are based on neural networks and combine information on
tracks and secondary vertices. The chosen working points correspond to a $\PQb$ tagging efficiency in
the range of 80--90\% and a mistagging rate around 10\%.
Reconstructed jets must not overlap with identified electrons,
muons, or \tauh within $\DR = \sqrt{\smash[b]{(\Delta\eta)^2+(\Delta\phi)^2}}<0.4$. To suppress jets
originating from instrumental background or from additional $\Pp\Pp$ interactions in the same and nearby bunch
crossings (pileup), a jet quality requirement based on the energy fraction of neutral hadrons and
charged hadrons associated with the primary vertex is applied~\cite{Khachatryan:2016kdb}.
The energy scale of jets is corrected for the nonlinear energy response of the calorimeters and the residual
differences between the jet energy scale in the data and in the simulation, separately in the different data taking periods.
The jet energy scale corrections are propagated to \ptmiss.

Leptons are required to originate from the primary vertex of the event to mitigate pileup effects.
An MVA classifier is used to distinguish between ``prompt'' electrons and muons coming
from $\PW$, $\cPZ$, or $\Pgt$ lepton decays
and ``nonprompt'' leptons originating from heavy-quark decays or  quark and gluon jets
incorrectly reconstructed as leptons. This MVA
classifier is trained using a set of observables related to the lepton kinematics, isolation, and
identification, as well as variables
relating the lepton to the nearest reconstructed PF jet, as described in Ref.~\cite{Sirunyan:2018shy}.
The requirement of this lepton MVA classifier, referred to as the ``tight'' selection,  corresponds to
a selection efficiency for  prompt leptons of about 90\%, and has an efficiency for nonprompt leptons at the percent level.
Further selection criteria are applied to ensure the correct
assignment of the electric charge in the reconstruction. These selection criteria include requirements
on the number of hits in the pixel system for electrons and on the agreement in the charge assignments of
multiple reconstruction algorithms for muons.

Events are selected using a combination of dilepton and single-lepton triggers with different lepton \pt thresholds.
The minimum \pt threshold requirements on the leading (subleading) lepton for the
electron-muon and dimuon triggers are 23\,(8) and 17\,(8)\GeV, respectively.
The single-lepton triggers, used to increase
the trigger efficiency, employ lepton \pt thresholds of 32 or 35\GeV for electrons and 24 or 27\GeV for muons.

The signal process is characterized by the presence of a pair of leptons of the same electric charge, along with a moderate
amount of \ptmiss originating from the neutrinos in the $\PW$ boson decays.

A ``loose'' set of requirements is imposed to retain a large set of events for training the  boosted
decision trees (BDT) that separate the signal from the main backgrounds~\cite{Hocker:2007ht}. Events are selected by requiring exactly two leptons of the same
charge, $\Pepm\PGmpm$ or $\PGmpm\PGmpm$, with \pt greater than 25\,(20)\GeV for the leading (subleading) lepton,
and $\abs{\eta}<2.5$ (2.4) for electrons (muons). Events are vetoed if there are additional leptons fulfilling
 looser identification and isolation requirements. The \pt thresholds for these additional leptons
are 7\GeV for electrons, 5\GeV for muons, and 20\GeV for \tauh candidates.
A lower threshold  of 15\GeV is applied to \ptmiss, which retains most of the signal
events, while significantly reducing the contributions from QCD multijet production, \ie, events from heavy- and
light-flavor jets
produced via strong interactions. The signal process involves no jet activity at LO although
around 25\% of signal events contain at least one reconstructed jet
with $\pt^{\text{jet}}>30\GeV$ within $\abs{\eta_{\text{jet}}}<2.5$. To ensure high signal efficiency, a requirement of at most one such jet is imposed.
Processes with $\PQb$ quark jets, such as \ttbar, are further suppressed
by rejecting events with at least one $\PQb$-tagged jet having $\pt^{\PQb\,\text{jet}}>25\GeV$ and $\abs{\eta_{\PQb\,\text{jet}}}<2.4$. The event selection criteria are summarized in Table~\ref{tab:evsel}.

\begin{table}[ht!]
\centering
\topcaption{Event selection criteria.}    \label{tab:evsel}
\begin{tabular}{l}
\hline
    Two leptons: $\Pepm\PGmpm$ or $\PGmpm\PGmpm$ \\
    $\pt^{\ell_1}>25\GeV$ , $\pt^{\ell_2}>20\GeV$ \\
    $\abs{\eta_{\Pe}}<2.5$, $\abs{\eta_{\PGm}}<2.4$ \\
    $\ptmiss>15\GeV$ \\
    $N_{\mathrm{jets}}<2$ ($\pt^{\text{jet}}>30\GeV$ and $\abs{\eta_{\text{jet}}}<2.5$) \\
    $N_{\PQb\mathrm{-tagged\,jets}}=0$ ($\pt^{\PQb\,\text{jet}}>25\GeV$ and  $\abs{\eta_{\PQb\,\text{jet}}}<2.4$) \\
    Veto on additional $\Pe$, $\PGm$, and \tauh candidates\\
\hline
\end{tabular}
\end{table}

\section{Simulated samples}\label{sec:dataMCsamples}

 A set of simulated samples is used to estimate the signal
and some of the backgrounds, whereas other backgrounds are estimated
using the data control regions, as described below.

 The signal process is simulated at LO in
perturbation theory using the \PYTHIA 8.226~\cite{Sjostrand:2014zea} event generator with the underlying tune
 CUETP8M1~\cite{sigeff7} for 2016, and  \PYTHIA 8.230 with the tune
  CP5~\cite{CMS-PAS-GEN-17-001} for 2017 conditions. The resulting values for $\sigma_{\text{eff}}$ of the two \PYTHIA tunes
are 29.9\unit{mb} for CUETP8M1 and 19.5\unit{mb} for CP5. The large difference between these values and the
resulting tune dependence
of $\sigma_{\mathrm{DPS\,\PW\PW}}$ underline the importance of measuring $\sigma_{\mathrm{DPS\,\PW\PW}}$ experimentally.
For the interpretation of the results a production cross section of 1.92 \unit{pb}, obtained with the CP5 tune, is used.

Another set of signal events is simulated using the MC event generator \HERWIGpp~\cite{Bahr:2008pv}
with tune CUETHppS1~\cite{sigeff7} and the CTEQ6L1~\cite{Pumplin:2002vw} PDF set. The kinematic observables are
described
consistently with the \PYTHIA
and \HERWIGpp event generators. Neither the underlying generator tune, nor the different PDF sets used to
generate the samples, impact the kinematic observables relevant to the analysis.

{\tolerance=8000
The $\PW\cPZ$ process is simulated at next-to-LO (NLO) with \POWHEG version 2.0~\cite{powheg,Alioli:2010xd}
and
\MGvATNLO 2.3.3~\cite{Alwall:2014hca}.
The former is used for the central prediction of this background,
while the latter is used for the study of systematic differences in kinematic distributions.
 The $\PW\Pgg$ and $\cPZ\Pgg$ samples, relevant to the $\Pepm\PGmpm$ final state,
 are generated with the \MGvATNLO event generator at NLO and LO, respectively.
To account correctly for parton multiplicities larger than one in the matrix element calculations,
the FxFx jet merging scheme~\cite{Frederix:2012ps}
is used for the NLO samples, while the MLM jet merging scheme~\cite{Alwall:2007fs} is used for the LO samples.
The background contributions arising from $\PW\Pgg^{*}$ and $\cPZ\cPZ$ production processes are simulated
 at NLO with the \POWHEG event generator, while \MGvATNLO is used to simulate the SHS $\PW\PW$ process.
\par}

The generators are interfaced with \PYTHIA to model parton showering and hadronization with the same underlying tunes used
for the signal generation.
 The NNPDF PDF sets with version 3.0~\cite{Ball:2014uwa} are used for 2016, while NNPDF v3.1~\cite{Ball:2017nwa}
PDF sets are used for 2017 conditions in the simulation of all processes.
The CMS detector response is modeled in the simulated events using \GEANTfour~\cite{Geant}, and are reconstructed with the
same algorithms used for the data.
Simulated events are weighted to reproduce the pileup distribution measured in the
data. The average pileup in data was 23 in 2016 and 32 in 2017.
The simulated MC events are scaled to correspond to the respective theoretical cross sections using the highest order
prediction available in each case~\cite{Campbell:2011bn,Binoth:2008kt}.

\section{Background estimation}\label{sec:backgrounds}

{\tolerance=300
Background processes can be separated into two categories.
The first category consists of processes with genuine same-charge lepton pairs from leptonic decays
 of bosons produced in the hard scattering.
These processes include first and foremost
the $\PW\cPZ$ process, in which both bosons decay leptonically and one of the
leptons from the $\PZ$ boson decay is either out of
detector acceptance or does not pass the identification criteria. Other such processes
include $\PW\Pgg^{*}$, $\PW\Pgg$,
$\PZ\Pgg$, and $\cPZ\cPZ$ production, as well as---to a lesser extent---the SHS $\PW^{\pm}\PW^{\pm}$ and $\PW\PW\PW$ processes.
Processes involving associated production of $\PW/\cPZ$ bosons and photons contribute
via asymmetric conversions of the photons into lepton pairs
inside the detector. All these background components
are estimated from MC simulation after applying scale factors to correct for residual differences between simulation and
data in the selection, reconstruction, and the modeling of the trigger. These scale factors are measured
using a ``tag-and-probe'' method~\cite{Sirunyan:2018shy}.
\par}

{\tolerance=800
The second category consists of two types of experimental backgrounds that resemble the production of prompt,
same-charge lepton pairs. The first type includes nonprompt lepton
backgrounds in which one or two of the selected leptons do not originate from the decay of a massive boson
from the hard scattering. This background component is dominated by $\PW${+}jets and QCD multijet events, with smaller
contributions from \ttbar production. The second type of experimental background arises from the misassignment
of the charge of an electron in the reconstruction and is dominated by $\cPZ\to\tau\tau$ when both $\tau$ leptons
decay leptonically to form an electron-muon pair.
\par}

Nonprompt leptons arise largely from leptonic heavy-flavor decays and from jets misidentified as leptons.
The main difference between a nonprompt and a prompt lepton is
the presence of larger hadronic activity around the lepton direction for the former.
This hadronic activity influences the lepton isolation and identification variables, and consequently the
lepton MVA classifier used for the selection of leptons. The selection criterion on this lepton MVA variable
is relaxed to define loose lepton selection criteria, and the leptons selected with this relaxed
MVA threshold are called ``loose'' leptons.

The lepton misidentification rate, which is defined as the probability of a
``loose'' nonprompt lepton to pass the ``tight'' lepton selection criteria, is estimated directly from the data in a
sample dominated by nonprompt leptons from QCD multijet and $\PW${+}jets processes~\cite{Sirunyan:2018shy}.
This control sample is constructed by requiring exactly
one ``loose'' lepton and at least one jet with $\Delta R{(\text{jet},\,\ell)}>1.0$ away from the lepton.
To suppress contributions of prompt leptons from electroweak processes,
an upper limit of 40\GeV is imposed on both \ptmiss and the transverse mass of the lepton
and \ptmiss. The transverse mass of two objects is defined as
\begin{linenomath}
\begin{equation}
    \mT(1,\,2) = \sqrt{2\pt^{(1)}\pt^{(2)}[1 - \cos{\Delta\phi(1,\,2)}]},
\end{equation}
\end{linenomath}
where $\Delta\phi(1,\,2)$ corresponds to the azimuthal angular difference between
the momenta of the two objects.

The residual contamination of prompt leptons in this control sample is subtracted using simulation.
The lepton misidentification rate is measured separately for electrons and muons as a function of the lepton \pt and $\abs{\eta}$.

To estimate the contribution of events with nonprompt leptons to the signal region,
another control sample of events is defined in the data. It is composed of events in which either one or both leptons
fail the lepton MVA selection criteria
but pass the ``loose'' selection, resulting in a sample of ``tight-loose''  and ``loose-loose'' lepton pairs.
These events are
reweighted as a function of the lepton misidentification
rates to obtain the estimated contribution from this background in the signal region.

Similar to the nonprompt lepton background, the probability
for the charge of a lepton to be incorrectly reconstructed is calculated and applied to the selected
opposite-charge dilepton events in data.
This lepton \pt-$\abs{\eta}$ dependent charge misidentification rate is measured in $\PZ\to\Pe\Pe$
events as the ratio of same-charge to opposite-charge dilepton events. Its value
ranges from 0.02\,(0.01)\% for electrons in the barrel to 0.40\,(0.16)\% for electrons in the endcaps for 2016\,(2017) data.
The charge misidentification rate for muons is negligible.

\section{Multivariate classifier training}
\label{sec:BDT}

The major background contributions arise from $\PW\cPZ$ production and processes with nonprompt leptons.
To separate the signal from these two background components, two separate
MVA classifiers are trained using a set of kinematic variables.

The $\PW\cPZ$ background is kinematically very similar to the signal, because they both have
two prompt leptons with moderate \ptmiss. Neither the signal nor the $\PW\cPZ$
process feature any hadronic activity in the form
of high-\pt jets at LO, and the masses
of the bosons decaying to leptons are very similar, resulting in similar \pt spectra for
the leptons. The main difference between the signal and $\PW\cPZ$ production is that in $\PW\cPZ$ production
the bosons share a Lorentz boost along the $z$-axis, whereas the bosons in the
signal process are approximately uncorrelated.

In the case of nonprompt lepton production, dominated by $\PW${+}jets and
QCD multijet processes, the kinematic differences with respect to the signal are larger.
However, these processes have production cross sections that are orders of magnitude larger than
that for the signal process. Therefore, even with a low probability of passing the event selection criteria, the
impact of these background processes is considerable.

A BDT-based framework combines this information to
 discriminate between the signal and the background events. The BDT training against the $\PW\cPZ$ sample is done
using its simulated sample, whereas the training against nonprompt leptons is carried out using a
 ``tight-loose'' control sample in data.

The following set of eleven input variables based on the lepton and event kinematics are used to train the BDTs:

\begin{itemize}
\item $\pt^{\ell_1}$ and $\pt^{\ell_2}$: transverse momenta of the two leptons;
\item \ptmiss;
\item $\eta_{\ell_1} \eta_{\ell_2}$:  product of pseudorapidities of the two leptons;
\item $\abs{\eta_{\ell_1}{+}\eta_{\ell_2}}$: absolute sum of $\eta$ of the two leptons;
\item $\mT(\ell_1,\,\ptmiss)$: transverse mass of the leading lepton and \ptmiss;
\item $\mT(\ell_1,\,\ell_2)$: transverse mass of the two leptons;
\item $\abs{\Delta\phi(\ell_1,\,\ell_2)}$: azimuthal angular separation between the leptons;
\item $\abs{\Delta\phi(\ell_2,\,\ptmiss)}$: azimuthal angular separation between the subleading lepton
      and \ptmiss;
\item $\abs{\Delta\phi(\ell_1\ell_2,\,\ell_2)}$: azimuthal angular separation between the dilepton system
      and the subleading lepton;
\item \mttwo: the so-called ``stransverse mass'' of the dilepton and \ptmiss system~\cite{mt21,mt22}.
\end{itemize}

The stransverse mass is defined as
\begin{linenomath}
\begin{equation}
\mttwo = \min\limits_{\ptvecmissO{+}\ptvecmissT=\ptvecmiss}\left[\mathrm{max}\left(\mT^{(1)}, \mT^{(2)}\right)\right],
\end{equation}
\end{linenomath}
in which \ptvecmiss is divided into two missing momentum
vectors, $\ptvecmissO$ and $\ptvecmissT$, to produce the transverse masses $\mT^{(1/2)}$ with the leptons in the event. In the case where both leptons
and both neutrinos originate from mother particles of equal mass, the $\mttwo$ variable exhibits an end point at the
mother particle mass.
All these variables show significant discrimination between the signal and background processes.
The background estimations describe the data well for these variables.

{\tolerance=500
 The two classifiers are mapped into a single
two-dimensional (2D) classifier by combining
contiguous regions in the 2D plane of the two separate classifiers.
These regions are chosen to optimize the constraining power of the maximum likelihood fit.
Namely, bins are chosen so that some exhibit large
signal-to-background ratio, while others are chosen to have small
signal contribution, but large contributions of either of the two main backgrounds. In total, the 2D plane is split into
15 such bins, on which the final fit is performed. Several different choices of mapping the 2D plane into a
one-dimensional (1D) classifier are tested according to these criteria, and the
 one exhibiting the largest expected significance for the signal is chosen.
\par}

Events are analyzed separately in the two distinct lepton flavor channels and the two---positive and
negative---charge configurations. Because the signal process is expected to be enhanced in the $\ell^+\ell^+$
configuration, while the background processes exhibit more symmetry between the two charges, the classification
into the two charge configurations increases the sensitivity of the analysis.

\section{Systematic uncertainties}\label{sec:systematics}

Various sources of systematic uncertainties, experimental and theoretical, can be grouped into
two categories. The first type changes the overall normalization
of one or more processes, whereas the second one can change both the normalization and the shape of the final 1D classifier distribution.
Their values and their correlation structure among the different data-taking periods and processes are described below.

The uncertainty in the integrated luminosity is 2.5 (2.3)\% for the 2016\,(2017) data-taking period~\cite{CMS-PAS-LUM-17-001,CMS-PAS-LUM-17-004}. The two
values are considered uncorrelated between the two years and are applied to all background processes estimated
from simulation, as well as the signal.

The dominant source of experimental systematic uncertainty is associated with the method adopted for the estimation of
nonprompt lepton contributions.
A normalization uncertainty of 40\,(25)\% for the $\Pepm\PGmpm$\,($\PGmpm\PGmpm$) final state is applied
to account for the observed variations in the performance of the background estimation method when applied to MC simulations.
Variations in the misidentification rate as a function of \pt and $\eta$ of the leptons are included  in addition to the
 uncertainties stemming from the kinematics of the event sample used to measure the lepton misidentification rate.
These kinematic
variations are estimated by varying the \pt of the jets in this sample, leading to
shape variations of the order of 5--10\% for the final classifier.
The overall normalization uncertainty is
considered correlated between the years, but uncorrelated between the two flavor channels, whereas the shape uncertainties
are considered fully uncorrelated between the years and flavor channels.

{\tolerance=300
A 30\% normalization uncertainty is applied to the
``charge misid.'' background in the
$\Pepm\PGmpm$ final state, covering the differences in the measurement of the
charge misidentification rate in data and simulation. This uncertainty is treated as fully correlated between the years.
\par}
Normalization uncertainties for the main backgrounds estimated from simulation are derived in dedicated
3 (4) lepton control regions
for the $\PW\cPZ$\,($\cPZ\cPZ$) processes. The scale factors
for the $\PW\cPZ$ and $\cPZ\cPZ$ processes are measured to be $1.01\pm 0.16$ and $0.97\pm 0.06$, respectively. The normalization uncertainties are estimated from the statistical
uncertainty and purity of these control samples and their scale factors, and take values of 16\,(6)\% for the $\PW\cPZ$\,($\cPZ\cPZ$) process. A 50\% normalization
uncertainty is applied to all other simulated backgrounds, accounting for the theoretical uncertainties in the predicted cross sections and
the lack of proper control samples in the data. The shape of the $\PW\cPZ$ process is allowed to
vary between the shapes coming from the two event
generators, \POWHEG and \MGvATNLO, and the corresponding uncertainty is considered correlated between 2016 and 2017.
The shape agreement in the kinematic observables between the
prediction and observation is of the order of 5\% in the $\PW\cPZ$ and $\cPZ\cPZ$ control regions.
It is assumed that the shape agreement is similar in all remaining simulation-derived backgrounds. Therefore a 5\% shape uncertainty
is applied to these components, which allows the shape of the final classifier to vary by up to 5\% linearly and quadratically
along the classifier distribution. The effect of variations in the renormalization and factorization scales
is negligible for the most important
background component, $\PW\cPZ$, and is therefore neglected.
Uncertainties in the PDF sets are expected to play a small role compared to the uncertainties described above.
Both the relevant generator level distributions such as the rapidity of the $\PW$ boson and the observable kinematic
variables used in the analysis are consistent between NNPDF sets v3.0 and v3.1.
An additional
complication in the estimation of the uncertainty in the PDFs arises because the standard procedures for evaluating
such uncertainties are ill-defined in the case of the signal process. For instance, varying a PDF set by any number of
replicas is an inadequate estimation of the modeling uncertainty
that emerges because the two PDF terms of the separate hard scatters are factorized when simulating signal events.
Therefore, such uncertainties are not considered. Rather, future measurements with
larger data sets will allow the study of the production cross section differentially in observables that are sensitive
to such nonfactorization effects.

The uncertainty in the pileup modeling is 1\% in the total yield for all simulated
backgrounds and the signal, and is assumed correlated among all flavors,
charges, and years. No significant differences in the kinematics
are observed because of the pileup modeling. The uncertainty in the $\PQb$ tagging
is considerably smaller than the statistical uncertainty in
the simulated samples and is therefore neglected. The acceptance
effect of the uncertainty in the jet energy scale is 2\% in the signal and the simulated background samples and is
considered fully correlated among all channels.

The trigger efficiency uncertainty associated with the combination of single-lepton and dilepton triggers is  1--2\%, whereas
the uncertainty in the data-to-simulation  scale factors for the ``loose''
lepton selection is 2\%. These uncertainties are considered correlated
among the flavor channels but uncorrelated between the years. The uncertainty in the ``tight'' lepton
selection is 2--3\%, and is considered correlated between the two years.

Any residual model dependence of the signal process is estimated by allowing the shape of the
DPS $\PW\PW$ process to vary between the \PYTHIA and \HERWIGpp simulations. The corresponding variations in
 the final BDT classifier are small.

Finally, the statistical uncertainty arising from the limited number of events in the simulated samples is included
independently for each bin of the final discriminant distribution for each final state and the two data-taking periods,
and is treated as fully
uncorrelated~\cite{Barlow:1993dm}.

\section{Statistical analysis and results}\label{sec:results}

Results are obtained after combining all the background and signal processes in the two separate flavor configurations,
$\Pe\Pgm$ and $\Pgm\Pgm$,
and two separate charge configurations, $\ell^+\ell^+$ and $\ell^-\ell^-$, resulting in four independent
distributions of the final BDT classifier. The final maximum likelihood
fit is performed simultaneously in these four distinct flavor and charge
categories~\cite{Khachatryan:2014jba,Cowan:2010js,CLS1}. The classification of
events into the two charge configurations increases the sensitivity of the analysis by 10\%.

Systematic uncertainties are represented in the likelihood by individual nuisance parameters,
and are profiled in the fit as described in Ref.~\cite{CMS-NOTE-2011-005}.
The number of events in each bin of the final classifier distribution used to extract the signal is modeled
as a Poisson random variable, with a mean value that is equal to the sum of signal and background contributions.

In total, 4921 events are observed in the four lepton-charge and flavor combinations. Table~\ref{tab:counts} summarizes
the yields of the various background and signal components along with their associated total uncertainties after
the ML fit (postfit).

\begin{table*}[ht!]
    \topcaption{Postfit background and signal yields and their uncertainties, and the observed event counts in the four charge and flavor combinations. The uncertainties include both statistical and systematic components. The SHS $\PW^{\pm}\PW^{\pm}$ and $\PW\PW\PW$ contributions are grouped as the ``Rare'' background.}
    \label{tab:counts}
\centering
      \begin{tabular}{ l c c c c c}
        \hline
                         & $\Pep\Pgmp$      & $\Pem\Pgmm$       & $\Pgmp\Pgmp$     & $\Pgmm\Pgmm$     \\\hline
Nonprompt                & $462\pm 71 $  & $411\pm 62  $  & $142\pm 31 $ & $118\pm 26 $\\
$\PW\cPZ$                & $834\pm 74 $  & $543\pm 50  $  & $537\pm 49 $ & $329\pm 31 $\\
$\cPZ\cPZ$               & $71\pm 6  $   & $66\pm 6   $   & $44\pm 4  $  & $38\pm 4  $\\
$\PW\Pgg^{*}$            & $256\pm 73 $  & $227\pm 65  $  & $133\pm 38 $ & $118\pm 34 $\\
Rare                     & $48\pm 17 $   & $23\pm 8   $   & $35\pm 13 $  & $14\pm 5  $\\
Charge misid.            & $17\pm 5  $   & $17\pm 5   $   & \NA              & \NA             \\
$\PW/\cPZ\Pgg$           & $131\pm 36 $  & $104\pm 28  $  & \NA              & \NA             \\ [\cmsTabSkip]
Total background         & $1819\pm 132$ & $1391\pm 107 $ & $891\pm 71 $ & $617\pm 53 $\\ [\cmsTabSkip]
DPS $\PW^{\pm}\PW^{\pm}$ & $77\pm 22 $   & $40\pm 12  $   & $57\pm 16 $  & $29\pm 9  $\\ [\cmsTabSkip]
Data                     & 1840             & 1480              & 926              & 675              \\
\hline
\end{tabular}
\end{table*}

\begin{figure*}[h!]
\centering
\includegraphics[width=0.425\textwidth]{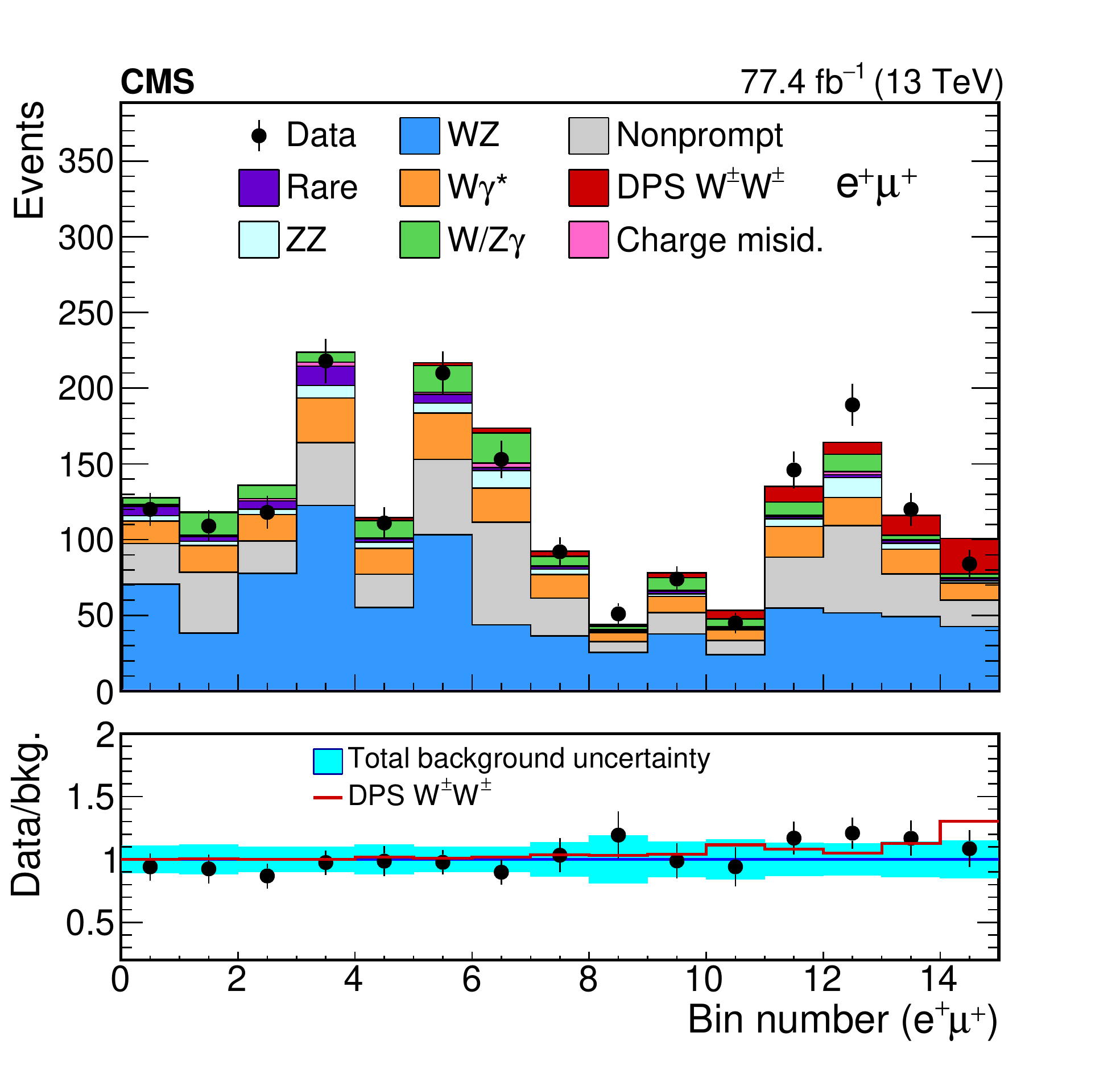}
\includegraphics[width=0.425\textwidth]{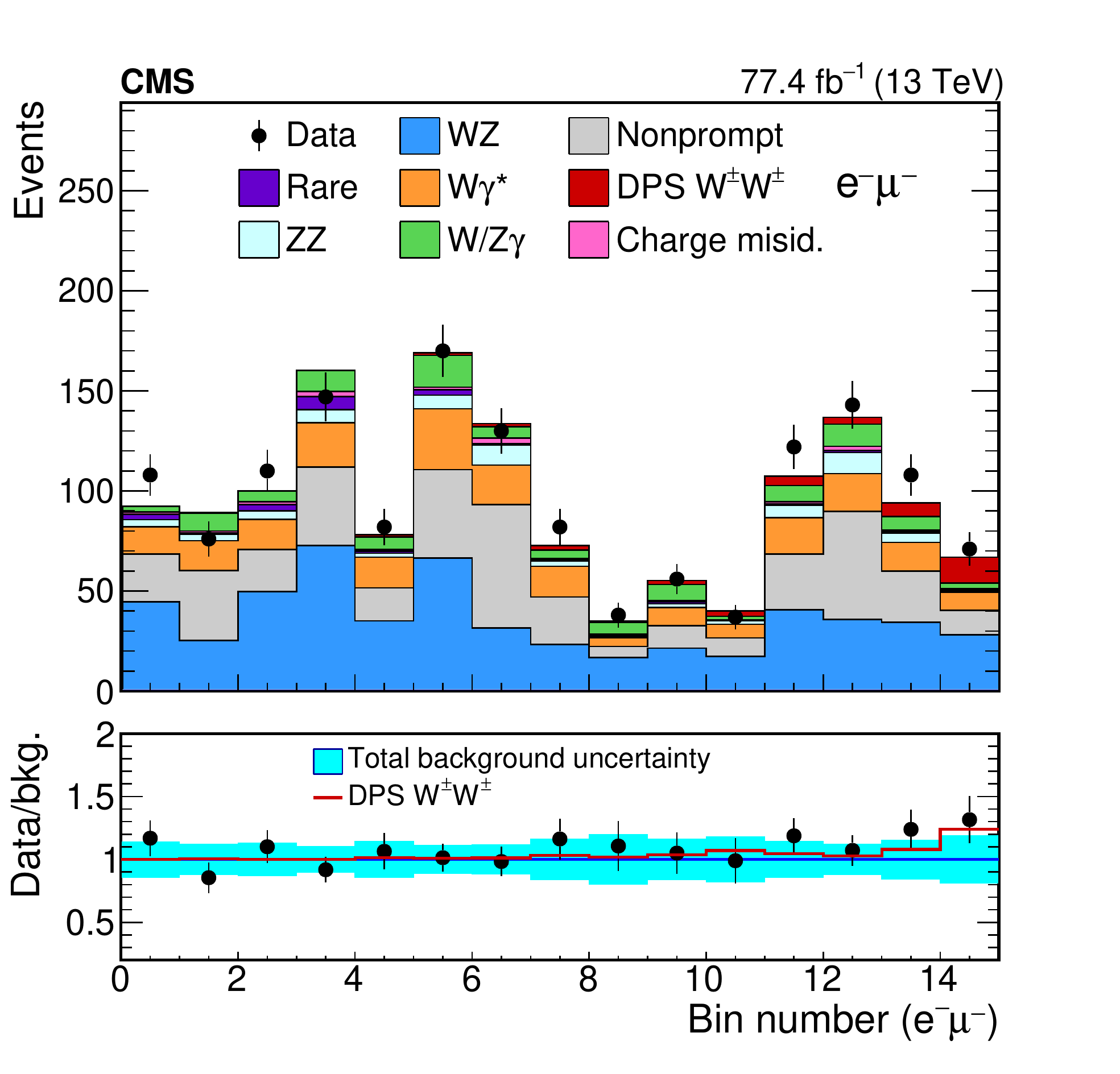}\\
\includegraphics[width=0.425\textwidth]{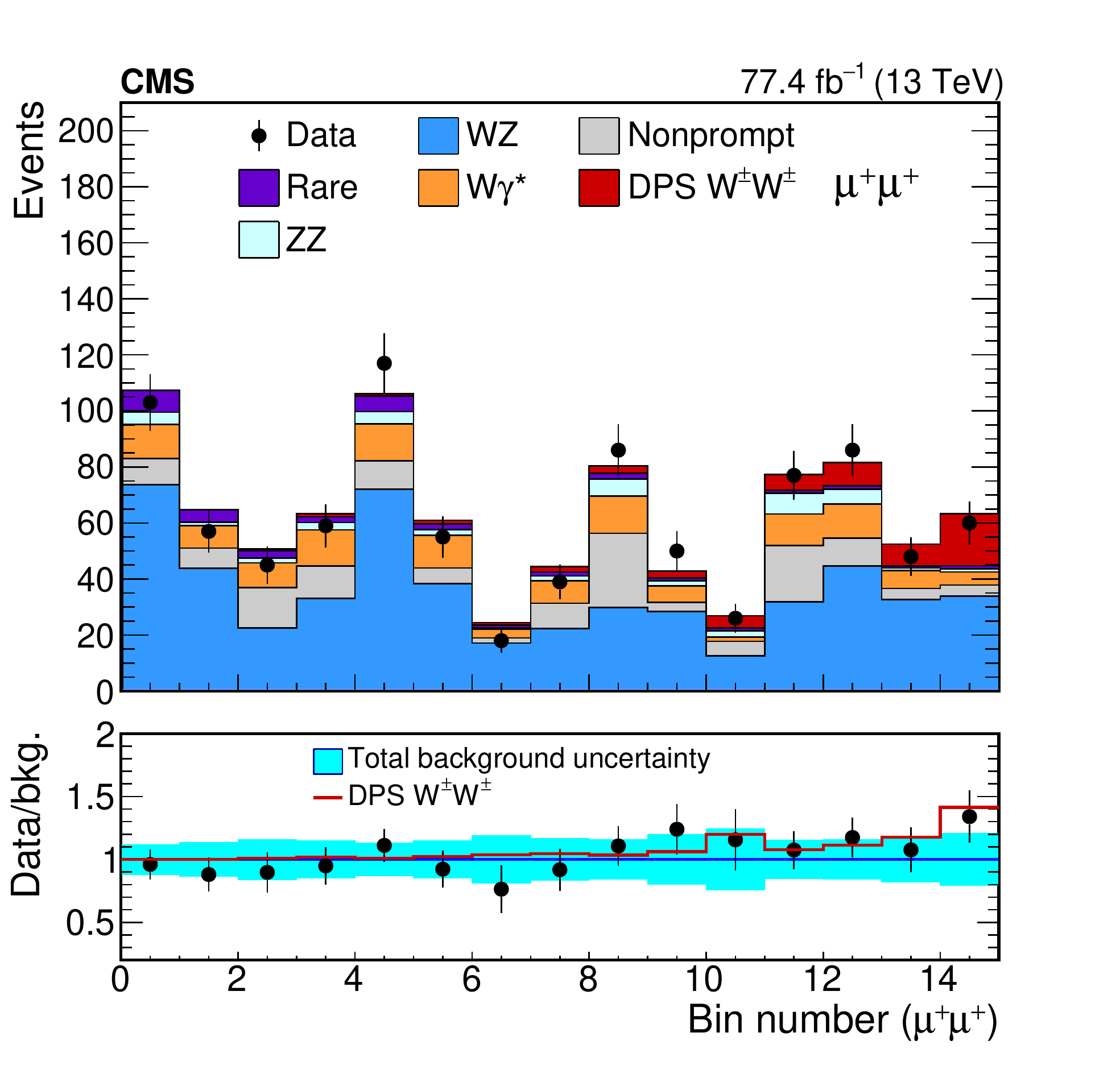}
\includegraphics[width=0.425\textwidth]{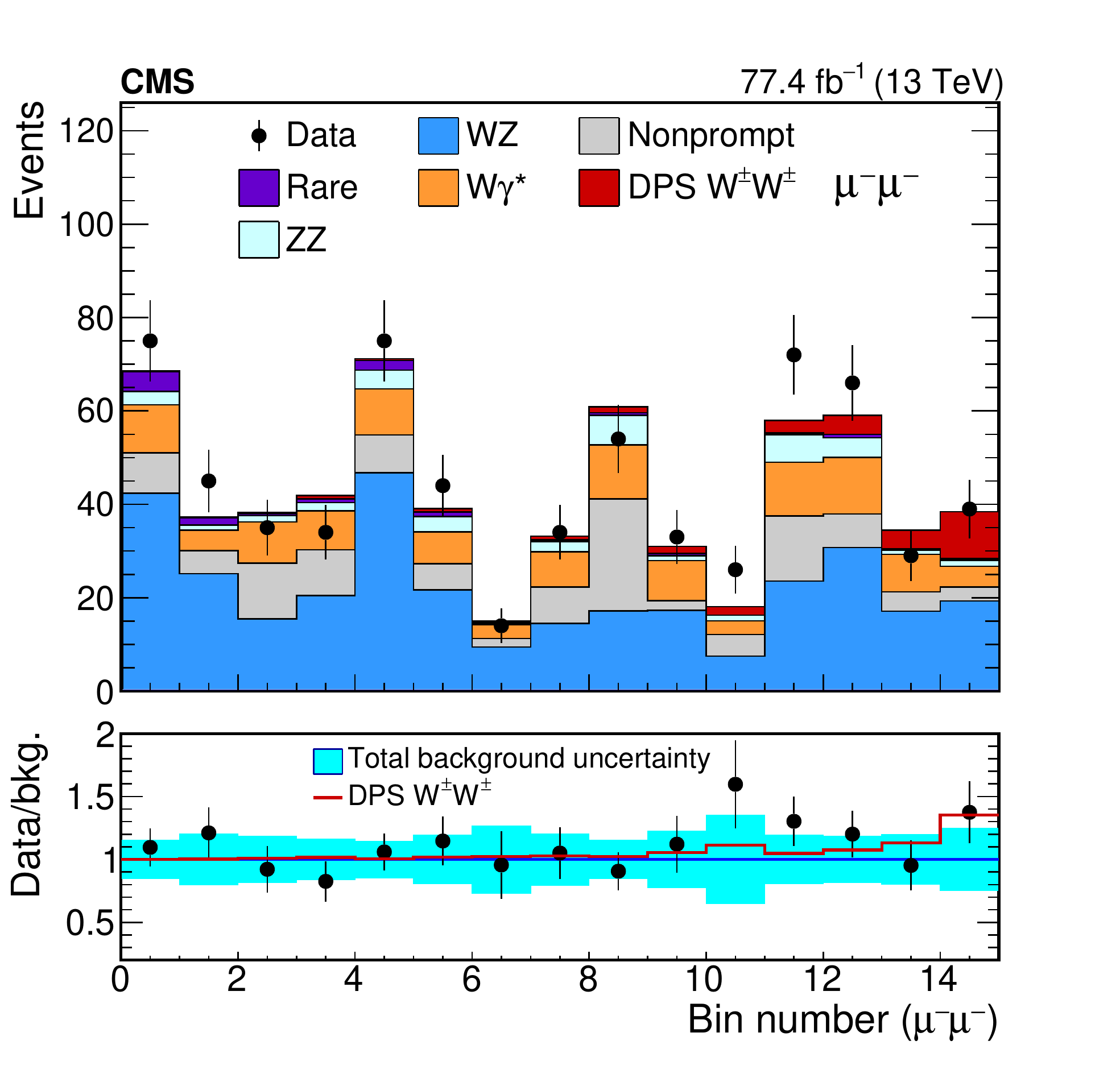}
\caption{Distribution of the final BDT classifier output for $\Pe\Pgm$ (upper) and $\Pgm\Pgm$ (lower) final states, in the
positive (left) and negative (right) charge configurations. Observed data are shown in black
markers while the backgrounds and signal are shown in colored histograms with their postfit yields. The SHS
$\PW^{\pm}\PW^{\pm}$ and $\PW\PW\PW$ contributions are grouped in the ``Rare'' background category.
The bottom panels show the ratio of data to the sum of all background contributions in the black
markers along with the signal shown using a red line.
The band represents the postfit background uncertainty, which includes both the statistical
and systematic components.
}\label{fig:results}
\end{figure*}

Figure~\ref{fig:results} shows the distribution of the final BDT classifier in the two charge
configurations in the $\Pe\Pgm$ channel
in the upper row,
and the two charge configurations in the $\Pgm\Pgm$ channel in the lower row, under the same scenario as in Table~\ref{tab:counts}, \ie,
postfit background and signal yields, together with the postfit total uncertainties.

Although the fit is performed with all the kinematic requirements applied, the following cross sections are
quoted as inclusive production cross sections for DPS $\PW\PW$.
The kinematic acceptance, defined as the ratio of events having a same-charge electron-muon or dimuon pair
from the $\PW$ boson decays and passing the analysis-level kinematic selection to the total number of generated events,
is measured using the \PYTHIA generator. In this definition, the leptons are used  at the ``dressed" level where the
momentum of a lepton is defined by combining its  pre-final-state radiation four-momentum with that of photons radiated
within a cone defined by $\DR = 0.1$ around the lepton. The kinematic acceptance is measured to be
$4.70\pm 0.02\stat\pm 0.94\model$\,\%.
The model uncertainty accounts for the differences in acceptance measured
using different PDF sets (NNPDF v3.0 and NNPDF v3.1), different \PYTHIA generator tunes (CUETP8M1 and CP5), and with
different event generators (\PYTHIA and \HERWIGpp). This uncertainty is dominated by the
differences seen between the \PYTHIA and \HERWIGpp event generators.

The prediction of any DPS $\PW\PW$ cross section suffers from large uncertainties. For the factorization
approach from Eq.~(\ref{eq:pocketformula}), the largest uncertainty comes from the imprecise knowledge of $\sigma_{\text{eff}}$,
which differs substantially between different measurements in different final states~\cite{sigeff1}.
Any predicted cross section from an
MC simulation, such as the one obtained from \PYTHIA also suffers from large uncertainties because of the tuning
of generator parameters sensitive to the modeling of the underlying event. Although the kinematic observables are tested
to be  unaffected by these tuning parameters, the predicted cross section varies by as much as 50\%. It is therefore
essential to interpret any ``predicted'' number in the following, either from the factorization approach or from \PYTHIA,
only as a rough estimate rather than a precisely derived quantity. Conversely, any observed cross section
and the corresponding significance do not depend on the predicted cross section, but only on the kinematics of the MC generator. These limitations emphasize
the importance of measuring the cross section of the DPS $\PW\PW$ process from data.

For this analysis, two predicted cross sections are used. The \PYTHIA event generator with the
CP5 tune gives a cross section of 1.92\unit{pb}. Alternatively, using Eq.~(\ref{eq:pocketformula})
 with the highest order cross section for inclusive $\PW$ boson production and decay at
 next-to-NLO accuracy in QCD and
NLO in electroweak corrections~\cite{Anastasiou:2003ds,fewz2p1}, $189\pm 7$\unit{nb}, along with
$\sigma_{\text{eff}}$ = $20.7\pm 6.6$\unit{mb}~\cite{sigeff1}, results in an expected cross section for the inclusive
DPS $\PW\PW$ process of $0.87\pm 0.28$\unit{pb}. The value
for $\sigma_{\text{eff}}$ is chosen as a representative number from a DPS cross section measurement based on a final state
containing a $\PW$ boson. A different choice of $\sigma_{\text{eff}}$ would alter the prediction of the cross section from the factorization approach accordingly.

The following quantities are obtained from the
simultaneous fit to the final BDT classifier in the four lepton charge and flavor combinations:

\begin{itemize}
    \item the expected significance assuming the signal
        process follows the \PYTHIA kinematics with the input cross section as $\sigma_{\mathrm{DPS\,\PW\PW,\,exp}}^{\PYTHIA}$;
    \item the expected significance assuming the signal process exhibits \PYTHIA-like kinematics with
      a production cross section, $\sigma_{\mathrm{DPS\,\PW\PW,\,exp}}^{\text{factorized}}$, extracted based on
      the factorization approach using the inclusive $\PW$ production cross section and
      value of $\sigma_{\text{eff}}$ mentioned above;
    \item the observed cross section $\sigma_{\mathrm{DPS\,\PW\PW,\,obs}}$ and the corresponding significance, assuming \PYTHIA-like kinematics, independent of the assumed cross section;
    \item $\sigma_{\text{eff}}$ using the inclusive $\PW$ production cross section and $\sigma_{\mathrm{DPS\,\PW\PW,\,obs}}$.
\end{itemize}

A maximum likelihood fit is performed separately for different lepton charge configurations and their
combination. The values obtained for the DPS $\PW^{\pm}\PW^{\pm}$ cross section are then extrapolated to the
inclusive $\PW\PW$ phase space.
Table~\ref{tab:expectedResults} summarizes the numbers extracted from the maximum likelihood fit
to the final classifier distribution for the combination of the $\ell^+\ell^+$ and $\ell^-\ell^-$ final states.

\begin{table*}[htb!]
    \topcaption{Results obtained from the maximum likelihood fit to the final classifier distribution. }
    \label{tab:expectedResults}
    \begin{center}
        \begin{tabular}{ l c c }
\hline
 & \multirow{2}{*}{Value}  & Significance \\
&  & (standard deviations)\\ \hline\noalign{\smallskip}
$\sigma_{\mathrm{DPS\,\PW\PW,\,exp}}^{\PYTHIA}$ & 1.92\unit{pb} & 5.4 \\ [\cmsTabSkip]
$\sigma_{\mathrm{DPS\,\PW\PW,\,exp}}^{\text{factorized}}$ & 0.87\unit{pb} & 2.5 \\ [\cmsTabSkip]
$\sigma_{\mathrm{DPS\,\PW\PW,\,obs}}$ & $1.41\pm 0.28\stat\pm 0.28\syst$\unit{pb} & 3.9 \\  [\cmsTabSkip]
$\sigma_{\text{eff}}$            & $12.7^{+ 5.0}_{- 2.9}$\unit{mb} & \NA \\ [\cmsTabSkip]
\hline

\end{tabular}
\end{center}
\end{table*}
{\tolerance=300
The observed inclusive DPS $\PW\PW$ production cross section is $1.41\pm 0.28\stat\pm 0.28\syst$\unit{pb} with
an observed significance of 3.9 standard deviations with respect to the background-only hypothesis.
This value lies between the prediction from \PYTHIA, which gives
a cross section of 1.92\unit{pb} with an expected significance of
5.4 standard deviations, and the one of the factorization approach, which predicts a cross section of 0.87\unit{pb} with an
expected significance of 2.5 standard deviations.
\par}

\begin{figure}[h!]
\centering
\includegraphics[width=0.48\textwidth]{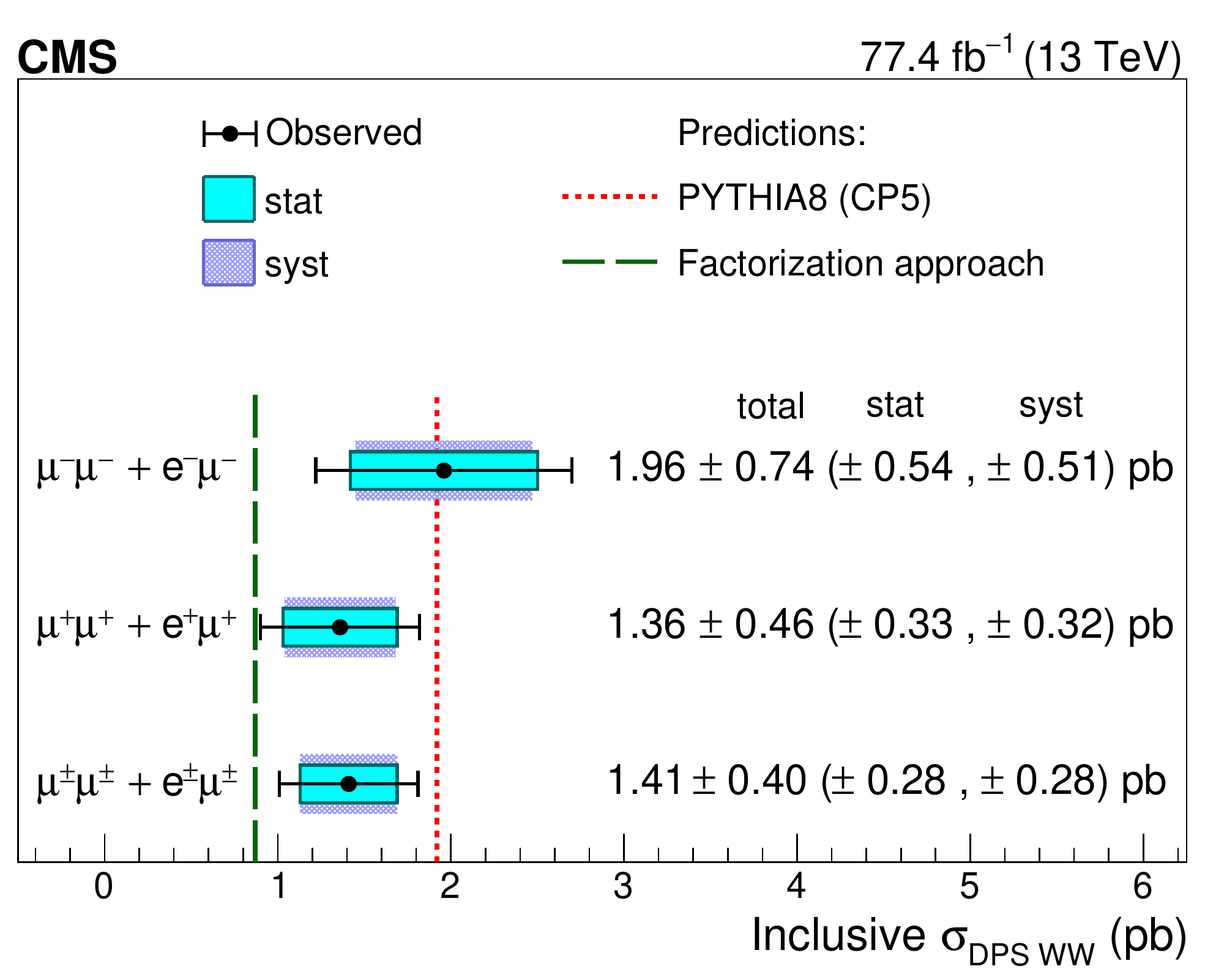}
\caption{Observed cross section values for inclusive DPS $\PW\PW$ production from the two
lepton charge configurations and their combination. These values are obtained from the extrapolation of
the observed DPS $\PW^{\pm}\PW^{\pm}$ cross section to the inclusive $\PW\PW$ case.
The statistical and systematic uncertainties are shown using shaded bands.
The predictions from \PYTHIA and from the factorization approach are represented
with the red dotted and green dashed lines, respectively.}

 \label{fig:xsecvalues}
 \end{figure}

The values of the inclusive DPS $\PW\PW$ production cross sections, obtained from the positive and negative lepton charge configurations,
 along with their combination, are shown in Fig.~\ref{fig:xsecvalues}.
The expected values for $\sigma_{\mathrm{DPS\,\PW\PW}}$, taken from \PYTHIA and the factorization approach, are also shown.
The positive charge configuration results in a measured inclusive cross section of $1.36\pm 0.33\stat\pm 0.32\syst$\unit{pb},
whereas for the negative charge configuration the value is $1.96\pm 0.54\stat\pm 0.51\syst$\unit{pb}.

A value of $\sigma_{\text{eff}}$ is extracted from Eq.~(\ref{eq:pocketformula}) in the following way. The SHS cross sections
for inclusive $\PW$ boson production are taken from theoretical calculations at next-to-NLO in QCD and
NLO in electroweak corrections, as described before. These cross sections are then combined with the
measured DPS $\PW^{\pm}\PW^{\pm}$ cross section, extrapolated to the full $\PW\PW$ phase space, to extract a
value for $\sigma_{\text{eff}}$.
This procedure results in a value for $\sigma_{\text{eff}}$ of
$12.7^{+ 5.0}_{- 2.9}$\unit{mb}, consistent with previous measurements of this quantity from
other final states~\cite{atlas:dps2018}.
This hybrid approach employed for calculating $\sigma_{\text{eff}}$ using, on the one hand, a theoretical prediction and, on the other hand,
the measured DPS $\PW\PW$ cross section results from the following consideration. Because
the statistical uncertainty dominates the measured $\sigma_{\mathrm{DPS\,\PW\PW}}$ and the leading systematic uncertainties are
specific to the $\ell^{\pm}\ell^{\pm}$ final state, these would not cancel with the uncertainties in a measurement
of the single $\PW$ boson production cross section. Therefore, the benefit of measuring the single $\PW$ boson production
cross section to extract a fully experimental value for $\sigma_{\text{eff}}$ is negligible.

\section{Summary}\label{sec:summary}
A study of $\PW\PW$ production from double-parton scattering (DPS) processes in proton-proton
collisions at $\sqrt{s}$ = 13\TeV has been reported. The analyzed data set corresponds to an integrated
luminosity of 77.4\fbinv, collected using the CMS detector in 2016 and 2017 at the LHC.
The $\PW\PW$ candidates are selected
in same-charge electron-muon or dimuon events with moderate missing transverse momentum and
low jet multiplicity.
Multivariate classifiers based on boosted decision trees are used to discriminate between
the signal and the dominant background processes.
A maximum likelihood fit is performed to extract the signal cross section, which is
compared to the predictions from simulation and from an approximate factorization approach.
A measurement of the DPS $\PW\PW$ cross section is achieved for the first time, and
a cross section of $1.41\pm 0.28\stat\pm 0.28\syst$\unit{pb} is extracted
with an observed significance of 3.9 standard deviations. This cross section leads to an effective cross section parameter of
$\sigma_{\text{eff}}=12.7^{+ 5.0}_{- 2.9}$\unit{mb}.
The results in this paper constitute the first evidence for $\PW\PW$ production from DPS.

\begin{acknowledgments}
We congratulate our colleagues in the CERN accelerator departments for the excellent performance of the LHC and thank the technical and administrative staffs at CERN and at other CMS institutes for their contributions to the success of the CMS effort. In addition, we gratefully acknowledge the computing centers and personnel of the Worldwide LHC Computing Grid for delivering so effectively the computing infrastructure essential to our analyses. Finally, we acknowledge the enduring support for the construction and operation of the LHC and the CMS detector provided by the following funding agencies: BMBWF and FWF (Austria); FNRS and FWO (Belgium); CNPq, CAPES, FAPERJ, FAPERGS, and FAPESP (Brazil); MES (Bulgaria); CERN; CAS, MoST, and NSFC (China); COLCIENCIAS (Colombia); MSES and CSF (Croatia); RPF (Cyprus); SENESCYT (Ecuador); MoER, ERC IUT, PUT and ERDF (Estonia); Academy of Finland, MEC, and HIP (Finland); CEA and CNRS/IN2P3 (France); BMBF, DFG, and HGF (Germany); GSRT (Greece); NKFIA (Hungary); DAE and DST (India); IPM (Iran); SFI (Ireland); INFN (Italy); MSIP and NRF (Republic of Korea); MES (Latvia); LAS (Lithuania); MOE and UM (Malaysia); BUAP, CINVESTAV, CONACYT, LNS, SEP, and UASLP-FAI (Mexico); MOS (Montenegro); MBIE (New Zealand); PAEC (Pakistan); MSHE and NSC (Poland); FCT (Portugal); JINR (Dubna); MON, RosAtom, RAS, RFBR, and NRC KI (Russia); MESTD (Serbia); SEIDI, CPAN, PCTI, and FEDER (Spain); MOSTR (Sri Lanka); Swiss Funding Agencies (Switzerland); MST (Taipei); ThEPCenter, IPST, STAR, and NSTDA (Thailand); TUBITAK and TAEK (Turkey); NASU and SFFR (Ukraine); STFC (United Kingdom); DOE and NSF (USA).

\hyphenation{Rachada-pisek} Individuals have received support from the Marie-Curie program and the European Research Council and Horizon 2020 Grant, contract Nos.\ 675440, 752730, and 765710 (European Union); the Leventis Foundation; the A.P.\ Sloan Foundation; the Alexander von Humboldt Foundation; the Belgian Federal Science Policy Office; the Fonds pour la Formation \`a la Recherche dans l'Industrie et dans l'Agriculture (FRIA-Belgium); the Agentschap voor Innovatie door Wetenschap en Technologie (IWT-Belgium); the F.R.S.-FNRS and FWO (Belgium) under the ``Excellence of Science -- EOS" -- be.h project n.\ 30820817; the Beijing Municipal Science \& Technology Commission, No. Z181100004218003; the Ministry of Education, Youth and Sports (MEYS) of the Czech Republic; the Lend\"ulet (``Momentum") Program and the J\'anos Bolyai Research Scholarship of the Hungarian Academy of Sciences, the New National Excellence Program \'UNKP, the NKFIA research grants 123842, 123959, 124845, 124850, 125105, 128713, 128786, and 129058 (Hungary); the Council of Science and Industrial Research, India; the HOMING PLUS program of the Foundation for Polish Science, cofinanced from European Union, Regional Development Fund, the Mobility Plus program of the Ministry of Science and Higher Education, the National Science Center (Poland), contracts Harmonia 2014/14/M/ST2/00428, Opus 2014/13/B/ST2/02543, 2014/15/B/ST2/03998, and 2015/19/B/ST2/02861, Sonata-bis 2012/07/E/ST2/01406; the National Priorities Research Program by Qatar National Research Fund; the Ministry of Science and Education, grant no. 3.2989.2017 (Russia); the Programa Estatal de Fomento de la Investigaci{\'o}n Cient{\'i}fica y T{\'e}cnica de Excelencia Mar\'{\i}a de Maeztu, grant MDM-2015-0509 and the Programa Severo Ochoa del Principado de Asturias; the Thalis and Aristeia programs cofinanced by EU-ESF and the Greek NSRF; the Rachadapisek Sompot Fund for Postdoctoral Fellowship, Chulalongkorn University and the Chulalongkorn Academic into Its 2nd Century Project Advancement Project (Thailand); the Nvidia Corporation; the Welch Foundation, contract C-1845; and the Weston Havens Foundation (USA).
\end{acknowledgments}

\bibliography{auto_generated}   

\cleardoublepage \appendix\section{The CMS Collaboration \label{app:collab}}\begin{sloppypar}\hyphenpenalty=5000\widowpenalty=500\clubpenalty=5000\vskip\cmsinstskip
\textbf{Yerevan Physics Institute, Yerevan, Armenia}\\*[0pt]
A.M.~Sirunyan$^{\textrm{\dag}}$, A.~Tumasyan
\vskip\cmsinstskip
\textbf{Institut f\"{u}r Hochenergiephysik, Wien, Austria}\\*[0pt]
W.~Adam, F.~Ambrogi, T.~Bergauer, J.~Brandstetter, M.~Dragicevic, J.~Er\"{o}, A.~Escalante~Del~Valle, M.~Flechl, R.~Fr\"{u}hwirth\cmsAuthorMark{1}, M.~Jeitler\cmsAuthorMark{1}, N.~Krammer, I.~Kr\"{a}tschmer, D.~Liko, T.~Madlener, I.~Mikulec, N.~Rad, J.~Schieck\cmsAuthorMark{1}, R.~Sch\"{o}fbeck, M.~Spanring, D.~Spitzbart, W.~Waltenberger, C.-E.~Wulz\cmsAuthorMark{1}, M.~Zarucki
% ...
\vskip\cmsinstskip
\dag: Deceased\\
1:  Also at Vienna University of Technology, Vienna, Austria\\
% ...
\end{sloppypar}
\end{document}